\documentclass[11pt]{article}
\usepackage{a4wide}
\usepackage{amssymb}
\usepackage{graphicx}
\usepackage{subfigure}
\usepackage{epsfig}
\usepackage[tbtags]{amsmath}
\usepackage{exscale}

\begin{document}

\setcounter{footnote}{0}

\newcommand{\lp}{\ell_{\mathrm P}}

\newcommand{\md}{{\mathrm{d}}}
\newcommand{\tr}{\mathop{\mathrm{tr}}}
\newcommand{\sgn}{\mathop{\mathrm{sgn}}}

\newcommand*{\R}{{\mathbb R}}
\newcommand*{\N}{{\mathbb N}}
\newcommand*{\Z}{{\mathbb Z}}
\newcommand*{\Q}{{\mathbb Q}}
\newcommand*{\C}{{\mathbb C}}

\newcommand{\kp}{\ensuremath{K_\varphi}}
\newcommand{\kx}{\ensuremath{K_x}}
\newcommand{\ex}{\ensuremath{E^x}}
\newcommand{\ep}{\ensuremath{E^\varphi}}
\newcommand{\gp}{\ensuremath{\Gamma_\varphi}}
\newcommand{\mr}{\ensuremath{\sqrt{\frac{2M}{x}}}}
\newcommand{\mrl}{\ensuremath{\sqrt{2Mx}}}
\newcommand{\be}{\begin{equation}}
\newcommand{\ee}{\end{equation}}
\newcommand{\bea}{\begin{eqnarray}}
\newcommand{\eea}{\end{eqnarray}}
\newcommand{\dif}{\mathrm{d}}

\newcommand{\kpb}{\ensuremath{\bar{K}_\varphi}}
\newcommand{\kxb}{\ensuremath{\bar{K}_x}}
\newcommand{\exb}{\ensuremath{\bar{E}^x}}
\newcommand{\epb}{\ensuremath{\bar{E}^\varphi}}
\newcommand{\gpb}{\ensuremath{\bar{\Gamma}_\varphi}}
\newcommand{\ab}{\bar{\alpha}}

\begin{center}
{\Large Inhomogeneities, loop quantum gravity corrections, constraint algebra and general covariance} \\
\vspace{1.5em}
Rakesh Tibrewala \footnote{e-mail address: {\tt rtibs@iisertvm.ac.in}}
\\
\vspace{1em}
Indian Institute of Science Education and Research,
CET Campus, Trivandrum 695016, India
\end{center}

\begin{abstract}
Loop quantum gravity corrections, in the presence of inhomogeneities, can lead to a deformed constraint algebra. Such a deformation implies that the effective theory is no longer generally covariant. As a consequence, the geometrical concepts used in the classical theory lose their meaning. In the present paper we propose a method, based on canonical transformation on the phase space of the spherically symmetric effective theory, to systematically recover the classical constraint algebra in the presence of the inverse triad corrections as well as in the presence of the holonomy corrections. We show, by way of explicit example, that this also leads to the recovery of general covariance of the theory in the presence of inverse triad corrections, implying that one can once again use the geometrical concepts to analyze the solutions in the presence of these quantum gravity corrections.  
\end{abstract}

\section{Introduction}
Canonical formulation of a generally covariant theory leads to the presence of constraints, which in turn are a reflection of the diffeomorphism invariance of the theory. For instance, the canonical formulation of general relativity leads to the presence of the, so called, Hamiltonian and the diffeomorphism constraints. The symmetry properties of the theory under general coordinate transformations are encoded in the hypersurface deformation algebra satisfied by the Hamiltonian and the diffeomorphism constraints \cite{dirac, teitelboim, hojman}. 

In any canonical theory of quantum gravity, therefore, one of the most important considerations is whether the algebra of quantum constraints closes or not? And if it does, whether the algebra retains its classical form or not? A negative answer to the first question would imply that the diffeomorphism invariance of the theory is anomalous and a negative answer to the second question would mean that, though the symmetry is not anomalous, it is deformed and its relation to the classical symmetries has to be understood.

Loop quantum gravity (LQG), being a canonical approach to quantum gravity, also has to face this question (see \cite{outside view} for a nice introduction to the problem and \cite{LaddhaLQGreloaded, MadhavanEuclideanGravity, dittrich} for different approaches to it). However, the full theory has not yet been constructed and these questions, therefore, cannot be answered at this stage. Under these circumstances it is useful to explore symmetry reduced models where it might be easier to examine the possible implications of the quantization with regard to the above stated questions. 

The simplest possible models are the cosmological (mini-superspace) models where the spacetime is homogeneous. However, because of the very nature of these models, the diffeomorphism constraint is identically zero and the features of the full constraint algebra cannot be explored within this class of models. 

In inhomogeneous models, for instance cosmological models with perturbative inhomogeneity or spherically symmetric models with radial inhomogeneity, not all components of the diffeomorphism constraint are zero. One therefore hopes that an understanding of the quantum theory for such models will throw some light on the questions related to the closure of quantum constraint algebra in the full theory. In this work we will focus on the spherically symmetric (midi-superspace) models.

Unfortunately, even for midi-superspace models, the quantum theory is not fully under control. However, certain well motivated considerations allow one to incorporate some of the generic features of the quantum theory in the classical constraints (inverse triad corrections and holonomy corrections in the context of LQG, for instance). It is expected that these `effective' constraints will capture some of the aspects of the quantum theory and there analysis can throw some light on issues related to the closure and/or possible deformation of the quantum constraint algebra.

Such an analysis has been carried out from different points of view \cite{modesto, viqar, GambiniPullin1, HusainCritical, GambiniPullin2}. However, in all these works, one or the other form of simplification is made so as to avoid dealing with the complicated issue of the constraint algebra (or the issue is altogether neglected). In \cite{modesto}, for example, simplification is achieved by working in the interior of the Schwarzschild black hole horizon, where the spacetime is homogeneous. The field theoretic difficulties are thus evaded and one can use techniques successful in loop quantum cosmology (LQC). On the other hand, in \cite{viqar, GambiniPullin1} a gauge choice is made from the beginning such that the constraint algebra becomes classical (while in \cite{GambiniPullin2} simplification was achieved by rescaling the Lagrange multipliers).

In a different line of attack in \cite{LTB2}, none of the above mentioned simplifications were made and the issue of constraint algebra was dealt with squarely. It was revealed that with the incorporation of certain quantum corrections the constraint algebra retained its classical form. However, for other kinds of quantum corrections the constraint algebra was deformed. (Similarly, in perturbative LQC it has been found that the constraint algebra is deformed \cite{MartinGolam, BarrauReview}.) 

In \cite{modifiedhorizon, einstein maxwell} it was explicitly shown by way of examples  that for corrections where the constraint algebra retained its classical form, general covariance continued to hold even though the dynamics of the theory was modified, whereas in the presence of a deformed constraint algebra this was no longer true. In other words, for deformed constraint algebra, solutions of constraints and equations of motion did not map to solutions of constraints and equations of motion under coordinate transformation, thus implying loss of general covariance. This in turn implied that basic geometric concepts like the spacetime metric and black hole horizon lost their significance (and meaning).

The above results seem consistent with the analysis of \cite{teitelboim}. There, Teitelboim derives the `structure constants' of the algebra of constraints by invoking the principle of path independence of the dynamical evolution. Since path independence of the dynamical evolution is a statement of general covariance of the theory, it seems plausible that for a deformed constraint algebra, the diffeomorphism invariance of the theory might be lost. 

In light of the above discussion, it seems worthwhile to investigate the issue of the 
deformation of the constraint algebra in the presence of quantum gravity corrections in more detail and to see whether there is a possibility to regain the 
classical form of the constraint algebra in a systematic way. In the present paper we propose one such method. The proposal, though made here for the case of LQG, should be relevant for any canonical theory of quantum gravity. 

The basic idea is suggested by Teitelboim's paper referred to above and is the following: once the quantum gravity inspired corrections are incorporated into the \emph{effective} Hamiltonian, the theory is analyzed classically on the original phase space. In terms of the original canonical coordinates the constraint algebra is deformed and general covariance is apparently lost. However, there exists the possibility that this deformation of the constraint algebra is only indicating that the original canonical coordinates are not the `right' coordinates to use once quantum corrections have been incorporated and a canonical transformation to a new set of phase space coordinates might restore the constraint algebra to the classical form and thus, in light of \cite{teitelboim}, also restore the general covariance. 

In this paper we show that this is indeed the case. We begin in the next section by quickly reviewing the classical theory (with spherical symmetry) written in terms of the Ashtekar variables, paying special attention to the constraints and their Poisson bracket algebra. We also briefly recapitulate the earlier results on the deformation of the constraint algebra in the presence of LQG corrections and the corresponding loss of general covariance. In section 3 we show that this deformation of the constraint algebra, in the presence of inverse triad corrections, is only a reflection of the fact that one pair of the original canonically conjugate variables is not the right set of coordinates to use on the phase space of the effective theory. We explicitly demonstrate that with a canonical transformation to a suitable pair of variables, the constraint algebra of the effective theory is rendered classical. A similar analysis is carried out in section 4 in the presence of the holonomy corrections where again it is shown that the classical constraint algebra can be restored after performing a suitable canonical transformation. In section 5 we give an example showing that restoration of the classical constraint algebra in the presence of the inverse triad corrections also leads to a regaining of the general covariance of the theory. We conclude in section 6 by discussing implications of the above proposal and possible future directions. Some of the details on the canonical transformation have been relegated to an appendix.  

\section{Classical and LQG corrected constraints and constraint algebra: A recap}

In this section we recall the earlier results on constraints and constraint algebra in terms of the $\mathfrak{su}(2)$ Ashtekar variables suitable for a loop quantization of spherically symmetric models. We will be brief and present only the important expressions, referring the reader to \cite{SphSymmstates, SphSymmHam} for details on the formulation of the problem in terms of Ashtekar variables and to \cite{modifiedhorizon, einstein maxwell} for a detailed account of deformed constraint algebra and subsequent loss of general covariance in the presence of quantum corrections.

In LQG, general relativity is recast in terms of densitized triads and $\mathfrak{su}(2)$ connection variables and, in the canonical formulation, this leads to the presence of the Gauss constraint in addition to the standard Hamiltonian and diffeomorphism constraints. In the presence of spherical symmetry, 
after solving the Gauss constraint, one is left with $(\kp, \ep)$ and $(\kx, \ex)$ as the canonically conjugate variables obeying the commutation relations
\be \label{commutation relations}
\{\kp(x), \ep(y)\}=G\delta(x,y) \quad  ,  \quad \{\kx(x), \ex(y)\}=2G\delta(x,y).
\ee
Here $(\kp, \kx)$ are the extrinsic curvature components (since the connection variables are eliminated in favor of the extrinsic curvature components) and $(\ep, \ex)$ are the components of the densitized triads. 

In terms of these variables the Arnowitt--Deser--Misner (ADM) metric is
\begin{equation} \label{adm metric}
{\rm d}s^2=-N^2{\rm d}t^2+\frac{\ep\,^2}{|E^x|}({\rm d}x+N^x{\rm
  d}t)^2+ |E^x|{\rm d}\Omega^2
\end{equation}
where $N(t,x)$ is the lapse function and $N^x(t,x)$ is the only non-zero component of the shift vector (here $t$ is the time coordinate and $x$ is the radial coordinate).

In vacuum, the Hamiltonian constraint $H[N]$ and the diffeomorphism constraint $D[N^{x}]$, respectively, are:
\be\label{classical hamiltonian constraint}
H[N]=-\frac{1}{2G}\int \md x\, N|\ex|^{-\frac{1}{2}}\bigg[\kp^2\ep+2\kp\kx\ex+(1-\gp^2)\ep+2\gp'\ex \bigg]\approx0,
\ee
\be\label{diffeomorphism constraint}
D[N^x]=\frac{1}{2G}\int \md x\,N^x\bigg[2\kp'\ep-\kx E^{x'}\bigg]\approx0,
\ee
where $\Gamma_{\phi}=-E^{x'}/2\ep$ and the prime $(')$ denotes derivative with respect to $x$.
These constraints satisfy the following algebra
\bea \label{classical constraint algebra}
\{D[N^{x}],D[M^{x}]\} &=& D[N^{x}M^{x'}-N^{x'}M^{x}], \nonumber \\
\{D[N^{x}],H[N]\} &=& H[N'N^{x}], \nonumber \\
\{H[N],H[M]\} &=& D[|\ex|(\ep)^{-2}(NM'-N'M)].
\eea
As indicated earlier, the constraint algebra satisfied by the Hamiltonian and the diffeomorphism constraint is a statement of the general covariance of the theory. The general covariance of the theory, obvious in the usual covariant formulation, implies that the solution of the constraints and the equations of motion (which are derived from these constraints) is mapped to other solution(s) under coordinate transformation(s). Since this is well known, we do not give explicit example to demonstrate this. 

\subsection*{LQG corrections}

As already mentioned in the Introduction, in the absence of the availability of full (loop) quantization of midi-superspace models, one tries to incorporate certain well motivated corrections in the classical Hamiltonian. The diffeomorphism constraint is left unmodified since in the full theory it is implemented by using group averaging. Two modifications which have received most attention in the literature are (i) the inverse triad corrections, and (ii) the holonomy corrections. 

\subsection*{(i) Inverse triad corrections}
The inverse triad corrections arise because of the presence of inverse powers of $\ex$ in the Hamiltonian \eqref{classical hamiltonian constraint}. In the quantum theory the spectrum of the corresponding operator is discrete containing zero \cite{RovelliSmolinDiscrete, AbhayLewandowskiArea} and therefore the inverse operator is not defined. Following Thiemann \cite{thiemannQSDI, thiemannQSDV}, one can nevertheless construct an inverse operator which has the correct classical limit but which has a very different behavior in the ultraviolet regime. 

For spherically symmetric models, the difference between the classical expression and the corresponding quantum version (obtained by computing $\langle T_{g,k,\mu}|\hat{H}|T_{g,k,\mu}\rangle$, where $|T_{g,k,\mu}\rangle$ denotes the spherically symmetric spin network basis) can be parametrized in terms of a scalar function $\alpha(\ex)$ \cite{LTB1}, such that 
\be \label{inverse triad correction}
\frac{1}{\ex}\rightarrow\frac{\alpha(\ex)}{\ex}
\ee
wherever inverse powers of $\ex$ appear in the Hamiltonian. Here the function $\alpha(\ex)$ is \cite{LTB1}
\begin{equation} \label{alpha}
\alpha(\ex) = 2\sqrt{\ex}\frac{\sqrt{|\ex+\gamma \lp^2/2|}-
\sqrt{|\ex-\gamma \lp^2/2|}}{\gamma \lp^2}
\end{equation}
where $\lp$ is the Planck length and $\gamma$ is the Barbero-Immirzi parameter. In the classical regime, $\ex\gg\gamma\lp^{2}/2$ and $\alpha(\ex)\rightarrow1$ giving the correct classical expression.

Due to the presence of quantization ambiguities \cite{QuantAmbiguity} there are more than one possibilities for the inverse triad corrected Hamiltonian since different powers of $\ex$ in the Hamiltonian can be corrected by different functions of $\alpha(\ex)$. In general, the inverse triad corrected Hamiltonian will have the form
\be \label{inverse triad corrected hamiltonian}
H_{I}[N]=-\frac{1}{2G}\int \md x\, N\bigg[\alpha|\ex|^{-\frac{1}{2}}\kp^2\ep+2\bar{\alpha}\kp\kx|\ex|^{\frac{1}{2}}+\alpha|\ex|^{-\frac{1}{2}}(1-\gp^2)\ep \\
+ 2\bar{\alpha}\gp'|\ex|^\frac{1}{2} \bigg],
\ee
where, presently, $\ab(\ex)$ is some function dependent on $\alpha(\ex)$ but not necessarily equal to it.

With the diffeomorphism constraint left unmodified, its Poisson bracket with itself retains its classical form. The other two Poisson brackets can receive non-trivial corrections and are
\bea \label{constraint algebra for inverse triad corrections}
\{D[N^{x}],H_{I}[N]\} &=& H_{I}[N'N^{x}], \nonumber \\
\{H_{I}[N],H_{I}[M]\} &=& D[\bar{\alpha}^{2}|\ex|(\ep)^{-2}(NM'-N'M)],
\eea
and we see that the last Poisson bracket is different from the corresponding classical expression, with a non-trivial dependence on the quantum correction function $\bar{\alpha}$. 

In earlier works \cite{LTB2, modifiedhorizon, einstein maxwell} two distinct cases were considered -- (a) $\bar{\alpha}=1$ and (b) $\bar{\alpha}=\alpha$. For case (a), the algebra after quantum corrections is the same as the classical algebra and one would believe (based on the theorem in \cite{teitelboim}) that the resulting theory would be generally covariant. This is indeed the case (see \cite{modifiedhorizon, einstein maxwell} for explicit examples) and, therefore, all the geometric notions like the metric and the horizon are meaningful. 

In case (b), where $\bar{\alpha}=\alpha$, the constraint algebra is deformed and it needs to be checked whether the resulting theory is generally covariant or not. A simple check can be made to see that the theory is not generally covariant. If we look for the analogue of the Schwarzschild solution (by solving the constraints and the equations of motion for the canonical variables) we formally get the `metric' \cite{modifiedhorizon}:
\begin{equation}\label{schwarzschild inverse triad version II}
\not\!{{\rm d}}s^{2}=-\alpha^{-2}\left(1-\frac{2M}{x}\right)\not\!{\rm d}t^{2}+
\left(1-\frac{2M}{x}\right)^{-1}\not\!{\rm d}x^{2}+x^{2}\not\!{\rm
  d}\Omega^{2} \,,
\end{equation}
The slash in $\not\!{{\rm d}}s^{2}$ is used to indicate that this is a formal construct and, as explained below, does not have the meaning of a metric (which is a covariant object). Assuming for the moment that \eqref{schwarzschild inverse triad version II} is a genuine metric, if we subject it to standard coordinate transformation to go to the Painlev\'e--Gullstrand like coordinate system, we get
\begin{equation} \label{painleve inverse triad version II}
\not\!{\rm d}s^{2}=-\alpha^{-2}\left(1-\frac{2M}{x}\right)\not\!{\rm d}T^{2}+
\alpha^{-2}\not\!{\rm d}x^{2}+2\alpha^{-2}\sqrt{\alpha^{2}-1+\frac{2M}{x}}
\not\!{\rm d}x\not\!{\rm d}T+x^{2}\not\!{\rm d}\Omega^{2}.
\end{equation}

If this theory is generally covariant then the above `metric', derived from \eqref{schwarzschild inverse triad version II} after a coordinate transformation, should also solve all the constraints and equations of motion. Reading off $N, N^{x}, \ep \, {\rm and} \, \ex$ from the above expression, it is easy to check that (after using two equations of motion to determine $\kp$ and $\kx$) it is not so. In other words, the object in \eqref{painleve inverse triad version II} is not a solution of the theory, implying that solutions (as in \eqref{schwarzschild inverse triad version II}) are not mapped to solutions under coordinate transformations. Thus, we conclude, that because of the deformed constraint algebra general covariance is lost. And since, by definition, metric is a generally covariant object, the formal constructs in \eqref{schwarzschild inverse triad version II} and \eqref{painleve inverse triad version II} cannot be interpreted as spacetime metrics.

Here we would like to mention that an \emph{adhoc} fix to this loss of general covariance was proposed in \cite{modifiedhorizon, einstein maxwell}. It was suggested by the form of the equations of motion and constraints and required one to replace $N\rightarrow\alpha N$, once all the equations of motion and constraints were solved. 
  
\subsection*{(ii) Holonomy corrections}
The holonomy corrections arise because, in the quantum theory, the operators corresponding to the connection components do not exist. Only the exponentiated versions of these components (integrated over suitable paths) are well defined. The effect of this feature of the quantum theory can be incorporated by replacing the connection components by their sinusoids such that in the classical regime one recovers the correct classical expression.

For spherical symmetry, there are two such components -- $\kp$ and $\kx$. The component $\kp$ is along the angular direction and on a given spherical orbit (i.e. a given value of the radial coordinate $x$) its value is a constant and we can make the replacement $\kp\rightarrow\sin(\delta\kp)/\delta$ in the classical Hamiltonian. Here $\delta$ is the path length (relative to the Planck length, say) used to evaluate the holonomy and corresponds to the underlying discreteness of the quantum theory. This replacement is expected to capture some aspects of the fact that only holonomies of connection components are well defined in the quantum theory. 

The $\kx$ component is the component along the radial direction and is more delicate to handle because of the presence of non-trivial diffeomorphisms along the radial direction. One cannot now use the simple point holonomy because of the inhomogeneity along the radial direction. For this reason we only discuss the corrections due to the $\kp$ holonomy, leaving the more difficult case of $\kx$ holonomy for future work.

In \eqref{classical hamiltonian constraint} we see that $\kp$ occurs at two places and that too with different powers. Therefore, in order to be general, as for inverse triad corrections, we include two independent correction functions $f_{1}(\kp)$ and $f_{2}(\kp)$ and fix the form of one relative to the other by demand of consistency. Explicitly, the Hamiltonian, after including the correction functions reads
\be \label{holonomy corrected hamiltonian}
H_{II}[N]=-\frac{1}{2G}\int \md x N|E^x|^{-\frac{1}{2}}(f_{1}^{2}E^\varphi+2f_{2}K_xE^x 
+(1-\gp^2)E^\varphi+2\gp'E^x).
\ee

With the diffeomorphism constraint left unmodified, it again turns out that only the Poisson bracket between two Hamiltonians is modified:
\be \label{hh bracket holonomy}
\{H_{II}[N],H_{II}[M]\}=D\bigg[\frac{\partial f_{2}}{\partial\kp}|\ex|(\ep)^{-2}(NM'-N'M)\bigg]+\frac{1}{2G}\int \md z(NM'-N'M)\frac{E^{x'}}{\ep}\left(f_{2}-f_{1}\frac{\partial f_{1}}{\partial\kp}\right).
\ee
Requirement of a first class constraint algebra implies that the second term on the rhs, which is unrelated to constraints, should be zero. This will be true generally if
\be \label{cond on holonomy corrections}
f_{2}-f_{1}\frac{\partial f_{1}}{\partial\kp}=0.
\ee

In accordance with the discussion above, we choose $f_{2}=\sin(\delta\kp)/\delta$ which leads to $f_{1}=2\sin(\delta\kp/2)/\delta$. Though, with the second term in \eqref{hh bracket holonomy} put to zero the constraint algebra has become first class, it is not the same as the classical algebra due to the presence of $\partial f_{2}/\partial\kp\equiv\cos(\delta\kp)$ in the first term on the rhs. Since the constraint algebra is deformed, we expect that general covariance must be lost. Since this has been demonstrated for deformed constraint algebra in case (b) with inverse triad corrections, we do not repeat the exercise for the present case. 

It is worth noting that unlike the case of inverse triad corrections where the classical algebra could be recovered by putting $\bar{\alpha}=1$, in the present case it is not possible to have the classical algebra without killing the holonomy corrections. This suggests that a deformed constraint algebra might be a generic feature of the quantum theory, thereby making an understanding (and implications) of such a deformation all the more important.  

\section{Inverse triad corrections and canonical transformation}
In this section we start by showing how, by performing a canonical transformation on the phase space with inverse triad corrected Hamiltonian, the constraint algebra in \eqref{constraint algebra for inverse triad corrections} can be rendered classical \eqref{classical constraint algebra}. Subsequently, in section 5, we give an explicit example showing that, as expected, a classical constraint algebra in the presence of quantum corrections leads to general covariance.

We begin by noting that the correction function $\bar{\alpha}^{2}$ in the Poisson bracket between two Hamiltonians in \eqref{constraint algebra for inverse triad corrections} depends only on $\ex$. Thus, if could absorb this function in a redefinition of $\ex$, there is a possibility that the Poisson bracket becomes classical. That is, we want a transformation from $\ex\rightarrow\exb$ such that
\be \label{relation between exbar and ex inverse triad}
\exb=\ab^{2}(\ex)\ex. 
\ee
(Note that the bar on $\ab(\ex)$ in the above equation has nothing to do with the bar on the new variable $\bar{\ex}$, the former being an inverse triad correction function as used in \eqref{inverse triad corrected hamiltonian}.)

To achieve this we perform a canonical transformation $(\kx, \ex)\rightarrow(\kxb, \exb)$, using the generating function $F_{3}\equiv F_{3}(\kxb,\ex)$, depending on the new coordinate $\bar{\kx}$ and the old momenta $\ex$ (and independent of the pair $(\kp, \ep)$). Specifically, we choose
\be \label{generating function for inverse triad corrections}
F_{3}(\kxb, \ex)=-\ab^{2}\ex\kxb.
\ee
This leads to 
\bea \label{canonical transformation for inverse triad correction}
\kx &=& -\frac{\partial F_{3}}{\partial\ex}=\bigg(\ab^{2}+2\ab\ex\frac{d\ab}{d\ex}\bigg)\kxb, \nonumber \\
\exb &=& -\frac{\partial F_{3}}{\partial\kxb}=\ab^{2}(\ex)\ex. 
\eea

Note that because $\ab(\ex)$ is a scalar, the density weights of $\kx$ and $\ex$ are not affected by this transformation. Also, since $\ab(\ex)$ is dimensionless, the dimensions of the new variables are the same as those of the old variables (as would be desirable if we are to interpret $\exb$ in the same way as $\ex$). Furthermore, in the classical regime $\ab(\ex)\rightarrow1$ and, as expected, the above transformation becomes the identity transformation.
 
Inverting the first equation in \eqref{canonical transformation for inverse triad correction} to express the new coordinate $\bar{\kx}$ in terms of the pair $(\kx, \ex)$, it is easy to verify that the new variables form a conjugate pair satisfying the commutation relation
\be \label{commutation relation for inverse triad corrections}
\{\kxb(x), \exb(y)\}=2G\delta(x,y)\,
\ee
with the other two Poisson brackets evaluating to zero (see appendix \ref{appendix inverse triad} for an explicit derivation). 

To complete the canonical transformation we need to express the diffeomorphism and the Hamiltonian constraints in terms of the new variables. For this we need to express the old variables $(\kx, \ex)$ entirely in terms of the new variables $(\kxb, \exb)$. We will assume that this can be done and proceed with writing the constraints in terms of the new variables (subsequently, when we work out an example, this step will be carried out explicitly). 

Before doing so, however, we also specialize to the case $\ab(\ex)=\alpha(\ex)$ since, as noted in the previous section, for this choice the algebra is deformed. Also, to avoid cluttering of notation, we continue to express the function $\alpha(\ex)$ by the same symbol $\alpha$, even when it is intended to be a function of $\exb$ and also drop the argument of $\alpha$ (we will follow a similar practice for other functions of old variables).

To express the diffeomorphism constraint, $\mathcal{D}=(2\kp'\ep-\kx E^{x'})/2G$ (in its unsmeared form), in terms of the new variables, we start by noting that (see equations \eqref{generating function for inverse triad corrections} and \eqref{canonical transformation for inverse triad correction})
\be 
\kx=\frac{d(\alpha^{2}\ex)}{d\ex}\kxb.
\ee
However, $\alpha^{2}\ex=\exb$, and therefore
\be
\kx=\frac{d\exb}{d\ex}\kxb=\kxb\left(\frac{d\ex}{d\exb}\right)^{-1},
\ee
where, in the expression on the right we assume that $\ex$ has been expressed as a function of $\exb$. Similarly, we can write
\be 
E^{x'}=\frac{d\ex}{d\exb}\bar{E}^{x'}.
\ee
Combining the two we have
\be 
\kx E^{x'}=\kxb\left(\frac{d\ex}{d\exb}\right)^{-1}\left(\frac{d\ex}{d\exb}\right)\bar{E}^{x'}=\kxb\bar{E}^{x'}.
\ee
Since the pair $(\kp, \ep)$ is left untouched, this implies that the diffeomorphism constraint retains its form in terms of the new variables. 

In a similar manner, by expressing the old variables in terms of the new variables and substituting them in the expression for the Hamiltonian \eqref{inverse triad corrected hamiltonian},  we find that the Hamiltonian constraint in terms of $(\kxb, \exb)$ is (with $d\alpha(\ex)/d\ex=b(\ex)$)
\bea \label{new hamiltonian inverse triad corrections}
\bar{H}_{I}[N] &=& -\frac{1}{2G}\int \md x\, N\bigg[\frac{\alpha^{2}\kp^{2}\ep}{|\exb|^{\frac{1}{2}}}+2\alpha^{2}\kp\kxb|\exb|^{\frac{1}{2}}+\frac{4\kp\kxb|\exb|^{\frac{3}{2}}b(\exb)}{\alpha}+\frac{\alpha^{2}\ep}{|\exb|^{\frac{1}{2}}} \nonumber \\
&&-\frac{(\bar{E}^{x'})^{2}}{4\alpha^{2}\ep|\exb|^{\frac{1}{2}}}-\frac{7(\alpha')^{2}|\exb|^{\frac{3}{2}}}{\alpha^{4}\ep}+\frac{5\alpha'|\exb|^{\frac{1}{2}}\bar{E}^{x'}}{\alpha^{3}\ep}-\frac{\bar{E}^{x''}|\exb|^{\frac{1}{2}}}{\alpha^{2}\ep}+\frac{2\alpha''|\exb|^{\frac{3}{2}}}{\alpha^{3}\ep} \nonumber \\
&&+\frac{E^{\varphi'}\bar{E}^{x'}|\exb|^{\frac{1}{2}}}{\alpha^{2}(\ep)^{2}}-\frac{2\alpha'E^{\varphi'}|\exb|^{\frac{3}{2}}}{\alpha^{3}(\ep)^{2}}\bigg] \approx 0.
\eea

Having written down the constraints in terms of the new variables, the next task is to check whether the constraint algebra acquires the classical form. From \eqref{alpha} we see that the form of the function $\alpha(\ex)$ is complicated and, therefore, inverting \eqref{canonical transformation for inverse triad correction} to obtain analytic expressions for $(\kx, \ex)$ in terms of $(\kxb, \exb)$ over the entire range of the radial coordinate $x$ could be difficult and one might have to do the inversion piece-wise. 

\begin{figure}
\begin{center}
\includegraphics{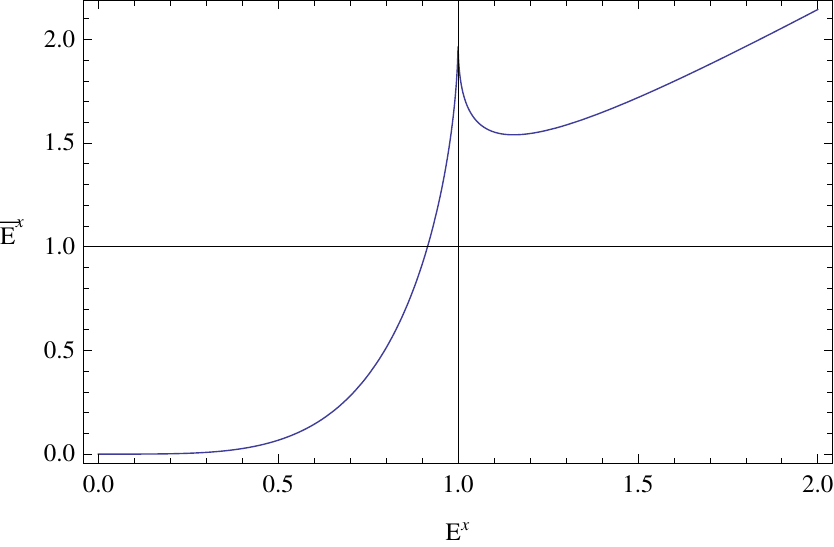}
\caption{\label{exbar vs ex} The new canonical variable $\exb=\bar{\alpha}(\ex)^{2}\ex$ as a function of the old canonical variable $\ex$. In the figure $\gamma \lp^2/2$ has been set equal to one.}
\end{center}
\end{figure}

In figure \ref{exbar vs ex} we plot equation \eqref{relation between exbar and ex inverse triad} showing $\exb$ as a function of $\ex$ (after setting $\gamma \lp^2/2=1$). We see that except for a small neighborhood around the peak of the curve at $\ex=\gamma \lp^2/2$, $\exb$ is a monotonic function of $\ex$ and therefore in these regimes \eqref{canonical transformation for inverse triad correction} can be inverted in a straight forward manner (it might still be difficult to obtain an analytic expression in which case one will have to resort to numerics). 

For $\ex\approx\gamma \lp^2/2$, $\exb(\ex)$ is not monotonic and three different domains on the $\ex$-axis correspond to a given neighborhood on the $\exb$-axis. For this regime therefore, the inversion has to be done piece-wise; that is, for each of these domains we use the formula appropriate for that domain (so that $\exb$ is monotonic within that domain) and then invert it. Accordingly, the constraints and the equations of motion will have to be solved piece-wise.

Since our primary aim is to demonstrate the recovery of the classical constraint algebra with the above transformations, we work in the regime $\ex\ll\gamma\lp^{2}/2$, where things turn out to be the simplest. In this regime \eqref{alpha} can be approximated by 
\be \label{approximate alpha}
\alpha(\ex)\approx\left(\frac{\gamma\lp^{2}}{2}\right)^{-3/2}(\ex)^{3/2}.
\ee
It is now easy to find
\bea \label{approximate can trans inv triad}
\ex &=& a_{\gamma}^{3/4}(\exb)^{1/4}, \nonumber \\
\kx &=& 4a_{\gamma}^{-3/4}(\exb)^{3/4}\kxb,
\eea
where, for the ease of notation, here and in the following we use $a_{\gamma}\equiv\gamma\lp^{2}/2$.

In this regime the Hamiltonian in \eqref{new hamiltonian inverse triad corrections} becomes
\bea \label{approx ham inv triad}
\bar{H}_{I}[N] &=& -\frac{1}{2G}\int \md x\, N\bigg[a_{\gamma}^{-3/4}\kp^{2}\ep(\exb)^{1/4}+8a_{\gamma}^{-3/4}\kp\kxb(\exb)^{5/4} \nonumber \\
&&+a_{\gamma}^{-3/4}\ep(\exb)^{1/4}+\frac{11}{64}a_{\gamma}^{3/4}\frac{(\exb)^{-5/4}(\bar{E}^{x'})^{2}}{\ep}-\frac{1}{4}a_{\gamma}^{3/4}\frac{(\exb)^{-1/4}\bar{E}^{x''}}{\ep} \nonumber \\
&&+\frac{1}{4}a_{\gamma}^{3/4}\frac{(\exb)^{-1/4}\bar{E}^{x'}E^{\varphi'}}{(\ep)^{2}}\bigg] \approx0. 
\eea
The diffeomorphism constraint, of course, has the form derived earlier, since that derivation was independent of the explicit form of $\alpha(\ex)$ and is
\be \label{approx diffeo inv triad}
\bar{D}[N^{x}]=\frac{1}{2G}\int \md x\, N^{x}(2\kp'\ep-\kxb\bar{E}^{x'})\approx0.
\ee 

We can now evaluate the Poisson bracket between the constraints. Since the diffeomorphism constraint retains its classical form, the Poisson bracket between two diffeomorphism constraints is unaffected and we only need to evaluate the bracket between the Hamiltonian and the diffeomorphism constraints and the bracket between two Hamiltonians. A straight forward exercise in algebra shows
\bea \label{approx constraint algebra inv triad}
\{\bar{H}_{I}[N], \bar{H}_{I}[M]\} &=& \bar{D}[\exb(\ep)^{-2}(NM'-N'M)], \nonumber \\
\{\bar{D}[N^{x}], \bar{H}_{I}[N]\} &=& \bar{H}_{I}[N'N^{x}].
\eea

As expected, the constraint algebra has acquired the classical structure after canonical transformation with $\ex$ replaced by $\exb$ in the first expression in \eqref{approx constraint algebra inv triad}. (From the derivation it should also be clear that the method would work irrespective of the functional form of $\alpha(\ex)$, the only requirement being that $\alpha$ is a function of $\ex$ alone.) We can now hope that this implies that in terms of the new variables the theory becomes generally covariant. In other words, the claim is that the metric, written in terms of $\exb$ in place of $\ex$, and given by the expression (compare with \eqref{adm metric})
\be \label{canonical transformed metric}
{\rm d}s^2=-N^2{\rm d}t^2+\frac{\ep\,^2}{|\bar{E}^{x}|}({\rm d}x+N^x{\rm
  d}t)^2+ |\bar{E}^{x}|{\rm d}\Omega^2,
\ee
is a covariant object in the presence of inverse triad corrections. In section 5 we show that this is indeed the case. However, before doing that, we show in the next section that a classical constraint algebra can be obtained even in the presence of holonomy corrections by a suitable canonical transformation.

\section{Holonomy corrections and canonical transformation}

In section 2, equation \eqref{hh bracket holonomy} we saw that in the presence of the holonomy corrections corresponding to $\kp$, the Poisson bracket between two Hamiltonians gives two terms, one of which is the diffeomorphism constraint and a second term which is unrelated to the constraints. As seen there, to have a first class algebra we needed to impose a condition which makes the second term zero. However, even after that the constraint algebra is deformed since
\be \label{constraint algebra holonomy}
\{H_{II}[N],H_{II}[M]\}=D[\cos(\delta\kp)\ex(\ep)^{-2}(NM'-N'M)],
\ee
and there is a factor of $\cos(\delta\kp)$ which is not present in the classical Poisson bracket. We now attempt to recover the classical constraint algebra for this class of corrections. (Note that the cosine function can take negative values; this corresponds to the signature change of the spacetime metric as noted in \cite{grain signature, martinpaily1}.)

Since the correction function $\cos(\delta\kp)$ on the rhs of \eqref{constraint algebra holonomy} depends on $\kp$, whereas classically there is no $\kp$ dependent term on the rhs, we try a canonical transformation from the set $(\kp, \ep) \rightarrow (\kpb, \epb)$ such that 
\[
\epb=\ep/|\cos(\delta\kp)|^{1/2},
\]
where the absolute value is used to take care of the negative values of the cosine function.

To achieve such a transformation, we choose a generating function $F_{2}\equiv F_{2}(\kp, \epb)$ depending on the old coordinate $\kp$ and the new momentum variable $\epb$ (and independent of the canonical pair $(\kx, \ex)$) with the explicit form
\be \label{generating function for holonomy correction}
F_{2}(\kp, \epb)=\epb\int |\cos(\delta\kp)|^{1/2}\md\kp.
\ee
Using this we find
\bea \label{canonical transformation holonomy}
\kpb &=& \frac{\partial F_{2}}{\partial\epb}=\int|\cos(\delta\kp)|^{1/2}\md\kp, \nonumber \\
\ep &=& \frac{\partial F_{2}}{\partial\kp}=\epb|\cos(\delta\kp)|^{1/2},
\eea
expressing the new coordinate $\kpb$ in terms of the old coordinate $\kp$ and giving the desired relation between the new momenta $\epb$ and the old pair $(\kp, \ep)$. Since $|\cos(\delta\kp)|$ is a scalar, the density weights of the transformed variables in the presence of holonomy corrections are the same as the original variables. Furthermore, as in the previous section, the dimensions of the new variables are the same as those of the old variables and in the classical limit, where $\delta\rightarrow0$, the above transformation gives $\kpb\rightarrow\kp$ and $\epb\rightarrow\ep$.

\begin{figure}
\begin{center}
\includegraphics[scale=0.9]{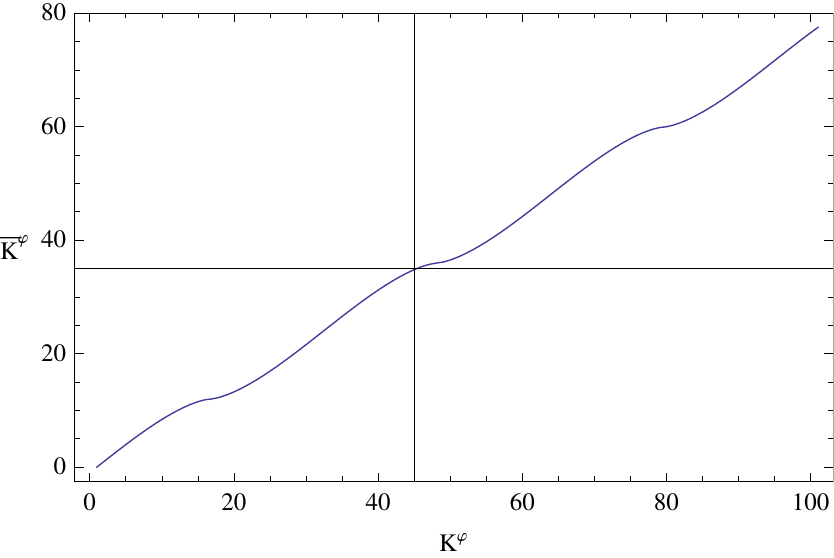}
\caption{\label{kpbar vs kp} The new canonical variable $\kpb=\int|\cos(\delta\kp)|^{1/2}\md\kp$ as a function of the old canonical variable $\kp$. In the figure $\delta=0.1$.}
\end{center}
\end{figure}

As in the previous section, to proceed further, we need to invert the expression relating $\kpb$ to the old variable $\kp$ in \eqref{canonical transformation holonomy}. In figure \ref{kpbar vs kp} we plot the first equation of \eqref{canonical transformation holonomy} showing the new variable $\kpb$ as a function of the old variable $\kp$ (with $\delta=0.1$). As seen, $\kpb$ is a monotonic function of $\kp$ and therefore, in principle, there is no problem in inverting the function in \eqref{canonical transformation holonomy} (though one might still have to resort to numerics).

We therefore assume that the first of the above two relations can be inverted to express $\kp$ in terms of $\kpb$ by a relation
\be \label{relation between kp and kpb}
\kp=s(\kpb).
\ee
Keeping the function $s(\kpb)$ unspecified also implies that this procedure will work for any deformation, the only condition being that the deformation factor in the $H$-$H$ Poisson bracket (see \eqref{constraint algebra holonomy}) is a function of $\kp$ alone.

Equation \eqref{relation between kp and kpb} implies that in the second equation in \eqref{canonical transformation holonomy} $\ep$ can be expressed entirely in terms of the new variables. It is now straight forward to check that the new coordinates satisfy the desired Poisson bracket relations, with the non-trivial bracket being
\be \label{commutation relations for holonomy corrections}
\{\kpb(x), \epb(y)\}=G\delta(x,y)
\ee
and the other two Poisson brackets vanishing. As for the case of inverse triad corrections, a derivation of these commutation relations is provided in section A.2 of the appendix.

Having expressed the old variables entirely in terms of the new variables, we now write the Hamiltonian constraint \eqref{holonomy corrected hamiltonian} (with $f_{1}=2\sin(\delta\kp/2)/\delta$ and $f_{2}=\sin(\delta\kp)/\delta$, see section 2) and the diffeomorphism constraint \eqref{diffeomorphism constraint} in terms of the new variables. Before doing so we note that because of \eqref{relation between kp and kpb}, we have $\cos(\delta\kp)=\cos(\delta s)$ and $\sin(\delta\kp)=\sin(\delta s)$ in terms of the new variable $\kpb$ (where, for the ease of notation, we ignore the argument in $s(\kpb)$) and therefore, $\ep=\epb(\cos(\delta s))^{1/2}$ (from here on we work in the regime where $\cos(\delta s)>0$ and thus ignore the absolute value sign; it is clear that the same method works for $\cos(\delta s)<0$ by replacing $\cos(\delta s)$ everywhere with $-\cos(\delta s)$). The Hamiltonian constraint can now be written as
\bea \label{new hamiltonian holonomy corrections}
\bar{H}_{II}[N] &=& -\frac{1}{2G}\int \md x\, N|E^{x}|^{-\frac{1}{2}}\bigg[\frac{4}{\delta^{2}}\sin^{2}\left(\frac{\delta s}{2}\right)(\cos(\delta s))^{1/2}\epb+\frac{2}{\delta}\sin(\delta s)\kx\ex \nonumber \\
&& +(\cos(\delta s))^{1/2}\epb-\frac{1}{4}\frac{(E^{x'})^{2}}{(\cos(\delta s))^{1/2}\epb}-\frac{\ex E^{x''}}{(\cos(\delta s))^{1/2}\epb} \nonumber \\ 
&& +\frac{\ex E^{x'}\bar{E}^{\varphi'}}{(\cos(\delta s))^{1/2}(\epb)^{2}}-\frac{\delta}{2}\frac{\ex E^{x'}\kpb'\sin(\delta s)}{(\cos(\delta s))^{3/2}\epb}\frac{ds}{d\kpb}\bigg].
\eea

To write the diffeomorphism constraint in terms of $(\kpb, \epb)$ we need to worry only for the $2\kp'\ep$ term in \eqref{diffeomorphism constraint}. To find what this is in terms of the new variables, we note that $\kp'=s'(\kpb)=\kpb'(ds/d\kpb)$ and therefore,
\[
2\kp'\ep=2\kpb'\frac{ds}{d\kpb}(\cos(\delta s))^{1/2}\epb.
\]
However, the first equation in \eqref{canonical transformation holonomy} implies that $d\kpb/d\kp=(\cos(\delta s))^{1/2}$ or, equivalently,
\[
\frac{d\kp}{d\kpb}\equiv\frac{ds}{d\kpb}=\frac{1}{(\cos(\delta s))^{1/2}}.
\]
Using this in the previous expression we find
\[
2\kp'\ep=2\kpb'\epb,
\]
implying that, just as for the case of inverse triad corrections, the form of the diffeomorphism constraint is unchanged when using canonically transformed variables in the presence of holonomy corrections:
\be \label{new diffeo holonomy}
D[N^{x}]=\frac{1}{2G}\int \md x N^{x}(2\kpb'\epb-\kx E^{x'}).
\ee

Proceeding analogously to the previous case of inverse triad corrections, we now need to evaluate the Poisson brackets between the constraints. With the Poisson bracket between two diffeomorphisms remaining unchanged, the other two Poisson brackets are found to be
\bea \label{new constraint algebra holonomy}
\{\bar{H}_{II}[N], \bar{H}_{II}[M]\} &=& D[\ex(\epb)^{-2}(NM'-N'M)], \\
\{D[N^{x}], \bar{H}_{II}[N]\} &=& \bar{H}_{II}[N'N^{x}],
\eea
and again we find that our intuition is justified and, in terms of the canonically transformed variables, we recover the classical constraint algebra in the presence of holonomy corrections as well. In light of the comment below \eqref{constraint algebra holonomy} regarding signature change, it is worth pointing that the canonical transformation presented here cannot undo the signature change. That is, in the regime where $\cos(\delta\kp)<0$, the surface deformation algebra is only brought to the classical form appropriate to the Euclidean signature.

\section{Regaining general covariance in the presence of inverse triad corrections: an example}
One of the implications of a deformed constraint algebra, as highlighted in \cite{modifiedhorizon, einstein maxwell}, is that the effective theory is not generally covariant. However, the arguments in \cite{teitelboim} suggest that a theory with classical constraint algebra should be generally covariant. We now verify whether this is the case in the presence of inverse triad corrections with the theory written in terms of the transformed variables which render the constraint algebra classical. 

The equations of motion, obtained using Hamilton's equations $\dot{A}=\{A, H[N]+D[N^{x}]\}$, with the Hamiltonian and the diffeomorphism given in \eqref{approx ham inv triad} and \eqref{approx diffeo inv triad}, respectively, are (here dot denotes derivative with respect to $t$):
\bea  
\dot{\bar{E}}^{x} &=& 8Na_{\gamma}^{-3/4}\kp(\exb)^{5/4}+N^{x}\bar{E}^{x'}, \label{exdot} \\
\dot{E}^{\varphi} &=& Na_{\gamma}^{-3/4}\kp\ep(\exb)^{1/4}+4Na_{\gamma}^{-3/4}\kxb(\exb)^{5/4}+N^{x'}\ep+N^{x}E^{\varphi'}, \label{ephidot} \\
\dot{\kp} &=& -\frac{N}{2}a_{\gamma}^{-3/4}\kp^{2}(\exb)^{1/4}-\frac{N}{2}a_{\gamma}^{-3/4}(\exb)^{1/4}+\frac{7N}{128}a_{\gamma}^{3/4}\frac{(\exb)^{-5/4}(\bar{E}^{x'})^{2}}{(\ep)^{2}} \nonumber \\
&&+\frac{N'}{8}a_{\gamma}^{3/4}\frac{(\exb)^{-1/4}\bar{E}^{x'}}{(\ep)^{2}}+N^{x}\kp', \label{kphidot} \\
\dot{\bar{K}}_{x} &=& -\frac{N}{4}a_{\gamma}^{-3/4}\kp^{2}\ep(\exb)^{-3/4}-10Na_{\gamma}^{-3/4}\kp\kxb(\exb)^{1/4}
-\frac{N}{4}a_{\gamma}^{-3/4}\ep(\exb)^{-3/4} \nonumber \\
&&+\frac{7N'}{32}a_{\gamma}^{3/4}\frac{(\exb)^{-5/4}\bar{E}^{x'}}{\ep}+\frac{7N}{32}a_{\gamma}^{3/4}\frac{(\exb)^{-5/4}\bar{E}^{x''}}{\ep}-\frac{7N}{32}a_{\gamma}^{3/4}\frac{(\exb)^{-5/4}E^{\varphi'}\bar{E}^{x'}}{(\ep)^{2}} \nonumber \\
&&+\frac{N''}{4}a_{\gamma}^{3/4}\frac{(\exb)^{-1/4}}{\ep}-\frac{N'}{4}a_{\gamma}^{3/4}\frac{(\exb)^{-1/4}E^{\varphi'}}{(\ep)^{2}}-\frac{35N}{256}a_{\gamma}^{3/4}\frac{(\exb)^{-9/4}(\bar{E}^{x'})^{2}}{\ep} \nonumber \\
&&+N^{x'}\kxb+N^{x}\kxb'. \label{kxdot}
\eea

In the classical regime, where $a_{\gamma}\ll\exb$, (static) solution to the corresponding set of equations would correspond to the Schwarzschild solution. However, since in the regime under consideration, $\exb\ll a_{\gamma}$ (see \eqref{approximate alpha}), a solution to the above set of equations would correspond to the corrections to the Schwarzschild solution in the deep quantum regime due to the inverse triad corrections.

We solve the above set of equations along with the Hamiltonian constraint \eqref{approx ham inv triad} and the diffeomorphism constraint \eqref{approx diffeo inv triad}, looking for a static solution. For this we assume $\kxb=\kp=N^{x}=0$ and also make the choice $\exb=x^{2}$ (implying that $4\pi x^{2}$ is the area of constant $t$ surfaces of radius $x$). 

With the above gauge choice, we see that \eqref{exdot} and \eqref{ephidot} along with the diffeomorphism constraint \eqref{approx diffeo inv triad} are already satisfied and only $\ep$ and $N$ are undetermined. We solve the Hamiltonian constraint, \eqref{approx ham inv triad}, for $\ep$ obtaining
\be \label{approx ephi inv triad}
\ep=\frac{a_{\gamma}^{3/4}}{\sqrt{16x+C_{1}a_{\gamma}^{3/2}x^{3/4}}}.
\ee 
This solution, when used in \eqref{kphidot}, gives
\be \label{equation for n'}
\frac{N'}{N}=\frac{2}{16x+C_{1}a_{\gamma}^{3/2}x^{3/4}}-\frac{7}{8x},
\ee
which, on integration, gives 
\be \label{approx n inv triad}
N=\frac{C_{2}\bigg[C_{1}a_{\gamma}^{3/2}+16x^{1/4}\bigg]^{1/2}}{x^{7/8}}.
\ee

In the above expressions, $C_{1}$ and $C_{2}$ are constants of integration which cannot be fixed in the present situation (since we are working in the deep quantum regime). These can only be fixed by first matching the solution to inverse triad corrected equations in the $\exb\gg\gamma\lp^{2}/2$ regime to the classical Schwarzschild solution and then using continuity arguments to match the solutions across $\exb=\gamma\lp^{2}/2$. In any case, as far as general covariance of the solution is concerned, the specific value of these constants is immaterial.

Although we have solved for $\ep$ and $N$, we have so far not used equation \eqref{kxdot} for $\kxb$ . This equation thus serves as a test for the gauge choice made. If the gauge choice is consistent, then this equation should be identically satisfied once the above solution along with the gauge choice is substituted on the rhs. It is straight forward to check that this is indeed the case implying that we have a consistent solution and we write the metric:
\be \label{approx schwarzschild metric inv triad}
ds^{2}=-\frac{C_{2}^{2}\left(16x^{1/4}+C_{1}a_{\gamma}^{3/2}\right)}{x^{7/4}}dt^{2}+\frac{a_{\gamma}^{3/2}}{x^{11/4}\left(16x^{1/4}+C_{1}a_{\gamma}^{3/2}\right)}dx^{2}+x^{2}d\Omega^{2}.
\ee

The most important test of the whole exercise, however, is that the above metric, under a coordinate transformation, should lead to a metric which is still a solution of the constraints and the equations of motion. For this, as in equations \eqref{schwarzschild inverse triad version II} and \eqref{painleve inverse triad version II}, we perform a coordinate transformation on the metric in \eqref{approx schwarzschild metric inv triad} from the Schwarzschild-like coordinates to Painlev\'e--Gullstrand-like coordinates.

This transformation makes use of the fact that in Schwarzschild-like coordinates, $\xi_{(t)}=\partial_{t}$ is a Killing vector (see \cite{modifiedhorizon} for details of the derivation). For a freely, in-falling radial geodesic with tangent $u^{a}\equiv(dt/dT, dx/dT, 0, 0)$ parameterized by proper time $T$, we have $g_{ab}u^{a}\xi^{b}_{(t)}=-1$ and $g_{ab}u^{a}u^{b}=-1$ (Latin indices $(a, b)$ taking values (0, 1)). Using these equations we find $u_{0}=-1$, and 
\be
u_{1}=-\frac{a_{\gamma}^{3/4}}{x^{11/8}\left(16x^{1/4}+C_{1}a_{\gamma}^{3/2}\right)^{1/2}}\left(\frac{x^{7/4}}{C_{2}^{2}\left(16x^{1/4}+C_{1}a_{\gamma}^{3/2}\right)}-1\right)^{1/2}.
\ee
Finally, using $-u_{a}dx^{a}=dT$ we can write
\be
dt=dT-\frac{a_{\gamma}^{3/4}}{x^{11/8}\left(16x^{1/4}+C_{1}a_{\gamma}^{3/2}\right)^{1/2}}\left(\frac{x^{7/4}}{C_{2}^{2}\left(16x^{1/4}+C_{1}a_{\gamma}^{3/2}\right)}-1\right)^{1/2}dx.
\ee
Substituting this in \eqref{approx schwarzschild metric inv triad}, we finally obtain the metric corresponding to Painlev\'e--Gullstrand like coordinates
\bea \label{approx painleve metric inv triad}
ds^{2} &=& -\frac{C_{2}^{2}\left(16x^{1/4}+C_{1}a_{\gamma}^{3/2}\right)}{x^{7/4}}dT^{2}
+\frac{C_{2}^{2}a_{\gamma}^{3/2}}{x^{9/2}}dx^{2} \nonumber \\
&&+2\left(\frac{C_{2}^{2}a_{\gamma}^{3/2}}{x^{9/2}}-\frac{C_{2}^{4}a_{\gamma}^{3/2}\left(16x^{1/4}+C_{1}a_{\gamma}^{3/2}\right)}{x^{25/4}}\right)^{1/2}dTdx+x^{2}d\Omega^{2}.
\eea

Comparing this with the general form of the ADM metric in \eqref{adm metric} we find $\exb=x^{2}$, $\ep=C_{2}a_{\gamma}^{3/4}/x^{5/4}$,
\be
N^{x}=\frac{x^{9/2}}{C_{2}^{2}a_{\gamma}^{3/2}}\bigg[\frac{C_{2}^{2}a_{\gamma}^{3/2}}{x^{9/2}}-\frac{C_{2}^{4}a_{\gamma}^{3/2}\left(16x^{1/4}+C_{1}a_{\gamma}^{3/2}\right)}{x^{25/4}}\bigg]^{1/2}
\ee
and, like for the classical solution in the Painlev\'e--Gullstrand coordinates, $N=1$. This solution, when used in \eqref{exdot} and \eqref{ephidot}, gives
\be
\kp=-\frac{x^{3}}{4C_{2}^{2}a_{\gamma}^{3/4}}\bigg[\frac{C_{2}^{2}a_{\gamma}^{3/2}}{x^{9/2}}-\frac{C_{2}^{4}a_{\gamma}^{3/2}\left(16x^{1/4}+C_{1}a_{\gamma}^{3/2}\right)}{x^{25/4}}\bigg]^{1/2}
\ee
and
\be 
\kxb=-\frac{C_{2}(6a_{\gamma}^{3/2}x^{7/4}+a_{\gamma}^{3}C_{1}C_{2}^{2})}{32x^{27/8}[a_{\gamma}^{3/2}C_{2}^{2}(x^{7/4}-16C_{2}^{2}x^{1/4}-a_{\gamma}^{3/2}C_{1}C_{2}^{2})]^{1/2}}.
\ee

If the theory is generally covariant then the above solution should satisfy all the remaining equations of motion and constraints, a highly non-trivial test of the overall consistency. It can be easily verified that this is indeed the case. Thus we have shown (though not proven) that in the presence of the inverse triad corrections, with suitable canonical transformation, the constraint algebra can be made classical which in turn makes the solution or the metric, written in terms of the transformed variables, generally covariant.

The next logical step would be to see whether the solution obtained by solving the constraints and the equations of motion in the presence of the $\kp$-holonomy corrections are generally covariant once the constraint algebra has been brought to the classical form as in \eqref{new constraint algebra holonomy}. However, the complicated nature of the transformations involved in equation \eqref{canonical transformation holonomy} makes the analysis difficult. Investigation of the general covariance of the solution in the presence of holonomy corrections is currently under progress.  

\section{Discussion and conclusions}

General covariance and the associated geometric notions are the basic properties of the classical theory of general relativity and much of the intuition about the solutions of the equations of the theory is based on them. One would therefore hope that these ideas would survive in the quantum version of the theory as well. Results based on LQC with perturbative inhomogeneities and those based on effective LQG corrections for spherically symmetric spacetimes suggest that general covariance is not automatic (in most cases) in the presence of LQG corrections, as indicated by the deformed algebra of constraints, and that the issue needs to be understood better.

In this paper we provided one systematic method for recovering the classical constraint algebra in the presence of inverse triad and holonomy corrections for spherically symmetric models. The method requires one to perform a suitable canonical transformation on the phase space of the effective theory such that the constraint algebra becomes classical. We also showed, by way of an example, that the recovery of the classical constraint algebra in the presence of the inverse triad corrections implied that the effective theory becomes generally covariant.

From the derivations presented, it should be apparent that in both the cases considered, the recovery of the classical constraint algebra does not depend on the specific form of the LQG correction function. It is also interesting to note that in both these cases, the diffeomorphism constraint retained its original form even after canonical transformation. 

We would now like to make a few comments and suggest some of the possible future directions. First point to note is that the present discussion has focused only on the inverse triad and the $\kp$-holonomy corrections. However, in the full quantum theory, several other corrections like the $\kx$-holonomy corrections and the higher moment corrections due to the quantum effects would be present. It needs to be seen whether the method works in these more general situations.

In this context it is worth while to note that even though the present work considered only the case of vacuum spacetime with spherical symmetry, the corrections to the constraint algebra (as in equations \eqref{constraint algebra for inverse triad corrections} and \eqref{constraint algebra holonomy}) are identical even in the presence of scalar matter \cite{modifiedhorizon} and $U(1)$ fields \cite{einstein maxwell}. We therefore expect the method proposed here to work in these more general cases as well.

Furthermore, instead of dealing with the inverse triad correction and the $\kp$-holonomy correction separately, as in this paper, we could consider these two corrections together. It then turns out that the deformation factor in the Poisson bracket of two Hamiltonians is simply the product of the deformation factor for the inverse triad correction and that for the holonomy correction. The canonical transformation would then involve the transformations of sections 3 and 4 simultaneously and the analysis should be straight forward.

Another question would be whether the method works only for spherically symmetric models or can it be extended to more complicated situations? This is a difficult question to answer as, so far, most of the applications of LQG have concerned with LQC (either homogeneous or with perturbative inhomogeneities) or with spherically symmetric models. 

For homogeneous LQC issues relating to the deformation of the constraint algebra do not arise while in perturbative LQC the algebra is not only deformed but also contains anomalous terms which are taken care of by introducing extra terms (counter-terms) to the Hamiltonian \cite{MartinAnomFreePertLQC}. It remains to be seen whether techniques similar to the present work can be fruitful to deal with anomalous constraint algebra and not just deformed constraint algebra.

Second point to note is that from the perspective of the quantum theory, working on the effective phase space is only an approximation. One would ideally like to work on the Hilbert space of the quantum theory. In that case, if it turns out that the algebra of quantum constraints is deformed like in the effective theory, the question to ask would be whether the proposal of the present work can be of any use there. This would be so if the canonical transformations, which absorb the deformation(s) of the constraint algebra, can be implemented unitarily on the Hilbert space of the quantum theory. 

It is also worth noting that in the case of holonomy corrections, the transformed triad $\epb$ depends on the old triad $\ep$ as well as the old curvature component $\kp$. This would imply that the metric, when written in terms of $\epb$, will effectively have dependence on the extrinsic curvature $\kp$. This suggests resemblance to proposals in gravity's rainbow \cite{LafranceMyers, AcciolyBlas, SmolinMagueijo}, where the metric is energy dependent and it needs to be explored whether there is something deeper in this analogy \footnote{Author is thankful to Martin Bojowald for pointing this out.}.    

As for the possible future directions, we already mentioned the ongoing work in the presence of the $\kp$-holonomy corrections regarding the recovery of general covariance in the presence of classical constraint algebra. Another possibility (as mentioned above) would be to see the usefulness of such methods in the presence of perturbative inhomogeneities in LQC. Since our main aim was to just present the basic ideas involved, we did not attempt to do a detailed analysis of the equations of motion and the resulting spacetime structure in terms of the transformed variables. 

In fact, because of the complicated equations involved, we did not look in the issue of general covariance in the presence of holonomy corrections. Even for inverse triad corrections we worked in the regime where the mathematical steps were explicit and did not require one to make further approximations once the form of the deformation function was fixed. In general, because of the complicated expressions involved, it would not be possible to find analytic expressions relating the old variables to the new and the use of numerical methods might be inevitable. We also briefly mentioned the fix provided in \cite{modifiedhorizon} to the problem of the loss of general covariance. Though the method was \emph{adhoc}, it still led to a generally covariant spacetime and it would be interesting to see how it compares with the present method. We plan to investigate some of these issues in the future.

We would like to end by emphasizing that one of the primary aims of the present work was to demonstrate that a deformed surface deformation algebra could be brought back to the classical form by performing canonical transformations on the phase space of the deformed theory. 

Since the constraint algebra encodes the symmetry properties of the theory, a deformed algebra suggests an apparent loss of general covariance. However, as demonstrated explicitly for the case of inverse triad correction, this could be an illusion, and the theory ccould become generally covariant once the constraint algebra is rendered classical. Also, as mentioned earlier, even though the present work focussed on corrections inspired by LQG, the method should be relevant for any canonical quantum theory of gravity where the constraint algebra gets deformed.

\section*{Acknowledgements}
The author would like to thank Martin Bojowald for useful discussions. He would also like to thank Prof. Hermann Nicolai for the invitation to and the hospitality at the Albert Einstein Institute, Golm (Potsdam), where a large part of this work was done. This work is supported under DST-Max Planck India Partner Group in Gravity and Cosmology.

\appendix

\section{Commutation relations for canonically transformed variables}
In this appendix we explicitly show that the new pair of variables -- $(\kxb, \exb)$ in the presence of inverse triad corrections and $(\kpb, \epb)$ in the presence of holonomy corrections -- are themselves conjugate and obey the correct Poisson bracket relations.

\subsection{Commutation relations for canonically transformed variables in the presence of the inverse triad corrections} \label{appendix inverse triad}

We start by showing that the pair $(\kxb, \exb)$ forms a conjugate pair and obeys the correct Poisson bracket relations. Referring back to equations \eqref{generating function for inverse triad corrections} and \eqref{canonical transformation for inverse triad correction}, we note that the new variables, when expressed in terms of the old variables are:
\be \label{new kx in terms of old kx}
\kxb=\frac{\kx}{\alpha^{2}+2\alpha\ex\frac{d\alpha}{d\ex}},
\ee
and
\be \label{new ex in terms of old ex}
\exb=(\alpha(\ex))^{2}\ex.
\ee

Since $\exb$ is independent of $\kx$, the Poisson bracket between two $\exb$'s obviously vanishes:
\be \label{ex-ex poisson bracket}
\{\exb(x), \exb(y)\}=\{\alpha^{2}\ex(x), \alpha^{2}\ex(y)\}=0.
\ee
Next, consider the bracket between two $\kxb$'s, which needs to be evaluated more carefully since, from \eqref{new kx in terms of old kx}, it depends on both $\kx$ and $\ex$. We have
\be \label{kx-kx poisson bracket}
\{\kxb(x), \kxb(y)\}=\bigg\{\frac{\kx(x)}{(\alpha^{2}+2\alpha\ex\frac{d\alpha}{d\ex})(x)} \, , \, \frac{\kx(y)}{(\alpha^{2}+2\alpha\ex\frac{d\alpha}{d\ex})(y)}\bigg\},
\ee
where, on the rhs, the notation $(F)(x)$ in the denominator implies that all the terms in $F$ are evaluated at $x$. We use the second relation in \eqref{commutation relations} to evaluate the relevant functional derivatives in the above expression to obtain
\bea
\{\kxb(x), \kxb(y)\} &=& 2G\int \md z\,\bigg[\frac{\delta(x,z)}{(\alpha^{2}+2\alpha\ex\frac{d\alpha}{d\ex})(x)}\frac{(-1)\kx(y)(4\alpha\frac{d\alpha}{d\ex}+2(\frac{d\alpha}{d\ex})^{2}\ex+2\alpha\frac{d^{2}\alpha}{d E^{x^{2}}}\ex)(y)\delta(y, z)}{(\alpha^{2}+2\alpha\ex\frac{d\alpha}{d\ex})^{2}(y)} \nonumber \\
&&-(x\leftrightarrow y) \bigg].
\eea
Using $\delta(x,z)$ to integrate with respect to $z$, we get
\be
\{\kxb(x), \kxb(y)\}=2G\bigg[\frac{-\kx(y)(4\alpha\frac{d\alpha}{d\ex}+2(\frac{d\alpha}{d\ex})^{2}\ex+2\alpha\frac{d^{2}\alpha}{d E^{x^{2}}}\ex)(y)}{(\alpha^{2}+2\alpha\ex\frac{d\alpha}{d\ex})(x)(\alpha^{2}+2\alpha\ex\frac{d\alpha}{d\ex})^{2}(y)} - (x\leftrightarrow y)\bigg]\delta(x, y),
\ee
which is zero because, for $x\neq y$, the $\delta$-function is zero and for $x=y$ the two terms in the square bracket are identical, giving zero on subtraction. We have thus shown that, as required, the Poisson bracket between two $\kxb$'s is zero.

Lastly, we evaluate the Poisson bracket between $\kxb$ and $\exb$
\bea
\{\kxb(x), \exb(y)\} &=& \bigg\{\frac{\kx(x)}{(\alpha^{2}+2\alpha\ex\frac{d\alpha}{d\ex})(x)} \, , \, \alpha^{2}\ex(y)\bigg\} \nonumber \\
&=& 2G\int \md z\,\bigg[\frac{\delta(x,z)}{(\alpha^{2}+2\alpha\ex\frac{d\alpha}{d\ex})(x)}(\alpha^{2}+2\alpha\ex\frac{d\alpha}{d\ex})(y)\delta(y,z)\bigg] \nonumber \\
&=& 2G\frac{(\alpha^{2}+2\alpha\ex\frac{d\alpha}{d\ex})(y)}{(\alpha^{2}+2\alpha\ex\frac{d\alpha}{d\ex})(x)}\delta(x,y).
\eea
When $x\neq y$, this correctly gives zero and for $x=y$, the numerator and the denominator cancel out, that is, the coefficient of the $\delta$-function is $2G$ as required (compare with the second equation of \eqref{commutation relations}).

\subsection{Commutation relations for canonically transformed variables in the presence of the holonomy corrections} \label{appendix holonomy}
To evaluate the Poisson bracket between the pair $(\kpb, \epb)$ we start by recalling the relations in equations \eqref{canonical transformation holonomy} and \eqref{relation between kp and kpb} to express the new variables in terms of the old (with $\cos(\delta\kp)>0$, see the remark above \eqref{new hamiltonian holonomy corrections})
\be \label{new kp in terms of old kp}
\kpb=\int(\cos(\delta\kp))^{1/2}\md\kp,
\ee
and
\be \label{new ep in terms of old kp and ep}
\epb=\ep(\cos(\delta\kp))^{-1/2}.
\ee
Since $\kpb$ in \eqref{new kp in terms of old kp} is independent of $\ep$, the Poisson bracket between two $\kpb$'s is trivially zero:
\be
\{\kpb(x), \kpb(y)\}=0.
\ee
Next consider the bracket between two $\epb$'s, which using the first relation in \eqref{commutation relations} is
\bea
\{\epb(x), \epb(y)\} &=& \{\ep(\cos(\delta\kp))^{-1/2}(x), \ep(\cos(\delta\kp))^{-1/2}(y)\} \nonumber \\
&=& G\int \md z\,\bigg[\frac{\delta\ep(x)\sin(\delta\kp(x))\delta(x,z)\delta(y,z)}{2(\cos(\delta\kp(x)))^{3/2}(\cos(\delta\kp(y)))^{1/2}}-(x\leftrightarrow y)\bigg] \nonumber \\
&=& G\bigg[\frac{\delta\ep(x)\sin(\delta\kp(x))}{2(\cos(\delta\kp(x)))^{3/2}(\cos(\delta\kp(y)))^{1/2}}-(x\leftrightarrow y)\bigg]\delta(x,y).
\eea
The last expression above implying that, as required, the Poisson bracket between two $\epb$'s is zero.

Finally, the Poisson bracket between $\kpb$ and $\epb$ is given by
\be
\{\kpb(x), \epb(y)\}=G\int \md z\bigg[\frac{\delta\kpb(x)}{\delta\kp(z)}\frac{\delta\epb(y)}{\delta\ep(z)}-\frac{\delta\kpb(x)}{\delta\ep(z)}\frac{\delta\epb(y)}{\delta\kp(z)}\bigg].
\ee
Since $\kpb$ is independent of $\ep$, the second term on the right in the above expression is zero and the first term, using equations \eqref{new kp in terms of old kp} and \eqref{new ep in terms of old kp and ep}, gives
\be 
\{\kpb(x), \epb(y)\}=G\bigg(\frac{\cos(\delta\kp(x))}{\cos(\delta\kp(y))}\bigg)^{1/2}\delta(x,y)\equiv\delta(x,y),
\ee
since for $x\neq y$ the $\delta$-function gives zero and for $x=y$ the coefficient of the $\delta$-function is unity.

\end{document}